\documentclass[aps,prb,twocolumn,showpacs,floatfix]{revtex4}
\usepackage[dvips]{graphics}
\usepackage{graphicx}
\usepackage[utf8]{inputenc}
\usepackage{amssymb}
\usepackage{epstopdf}
\usepackage{bm}

\begin{document}
\title{Josephson effect and spin-triplet pairing correlations in
SF$_1$F$_2$S junctions}
\author{Luka Trifunovic}
\altaffiliation[Present address: ]{Department of Physics, University of Basel,
Klingelbergstrasse 82, CH-4056 Basel, Switzerland}
\affiliation{Department of Physics, University of Belgrade, P.O. Box 368, 11001
Belgrade, Serbia}
\author{Zorica Popovi\'c}
\affiliation{Department of Physics, University of Belgrade, P.O. Box 368, 11001
Belgrade, Serbia}
\author{Zoran Radovi\'c}
\affiliation{Department of Physics, University of Belgrade, P.O. Box 368, 11001
Belgrade, Serbia}

\begin{abstract}
  We study theoretically the Josephson effect and pairing correlations in planar
  SF$_1$F$_2$S junctions that consist of conventional superconductors (S) connected
  by two metallic monodomain ferromagnets (F$_1$ and F$_2$) with transparent
  interfaces. We obtain both spin-singlet and -triplet pair amplitudes and
  the Josephson current-phase relations for arbitrary orientation of the
  magnetizations using the self-consistent solutions of Eilenberger equations in
  the clean limit and for a moderate disorder in ferromagnets. We find that the long-range
  spin-triplet correlations cannot prevail in symmetric junctions with equal
  ferromagnetic layers. Surprisingly, the long-range spin-triplet correlations
  give the dominant second harmonic in the Josephson current-phase relation of
  highly asymmetric SF$_1$F$_2$S junctions.
  The effect is robust against moderate disorder and variations in the layers
  thickness and exchange energy of ferromagnets.
\end{abstract}


\maketitle

\section{Introduction}
Spin-triplet superconducting correlations induced in heterostructures comprised
of superconductors with the usual spin-singlet pairing and inhomogeneous
ferromagnets have attracted considerable attention recently.~\cite{bergeret_odd_2005}
Triplet pairing which is odd in frequency was envisaged a long
ago~\cite{berezinskii__1974} in an attempt to describe the A phase of superfluid
$^3\rm{He}$. Even though it was found that the pairing in superfluid
$^3\rm{He}$ is odd in space (p-wave) rather than in time, it is predicted that
even in space (s-wave) and odd in time (odd in frequency) pairing does occur in certain superconductor
(S) -- ferromagnet (F) structures with inhomogeneous
magnetization.~\cite{bergeret_long-range_2001,volkov_odd_2003}
As a result, superconducting correlations can have a long-range
propagation from SF interfaces, with
penetration lengths up to 1$\mu m$ and a nonvanishing Josephson supercurrent
through very strong
ferromagnets.~\cite{sosnin_superconducting_2006,keizer_spin_2006,anwar_supercurrents_2010,wang_interplay_2010,
sprungmann_evidence_2010,khaire_observation_2010,robinson_controlled_2010}
%robinson_enhanced_2010

The first evidence of long-range S--F proximity effect came from the experiments on
long wires made of Ho conical ferromagnet,~\cite{sosnin_superconducting_2006}
and a fully-spin-polarized CrO$_2$ halfmetallic
ferromagnet,~\cite{keizer_spin_2006} in which Josephson supercurrent was
measured. These experimental findings were subsequently verified using different
substrates and superconducting contacts.~\cite{anwar_supercurrents_2010} Very
recently, the effect has also been observed in single-crystal ferromagnetic Co
nanowires,~\cite{wang_interplay_2010} a Heusler alloy
Cu$_2$MnAl,~\cite{sprungmann_evidence_2010} and synthetic antiferromagnets with
no net magnetization where thin PdNi or Ho layers are combined with Co
layers.~\cite{khaire_observation_2010,robinson_controlled_2010}
%Substantially
%enhanced supercurrent is also observed in Josephson junctions with ferromagnetic
%Fe/Cr/Fe trilayer in the antiparallel
%configuration, without long-range spin-triplet
%correlations.~\cite{robinson_enhanced_2010}

In SFS Josephson junctions with
homogeneous~\cite{buzdin_proximity_2005} or
spiral~\cite{kuli_possibility_2001} magnetization, the projection
of the total spin of a pair to the direction of magnetization is
conserved and only spin-singlet $f_s$ and triplet $f_{t0}$ correlations with
zero spin projection occur. These
correlations penetrate into the ferromagnet over a short distance determined
by the exchange energy.
%, except for clean single-channel junctions.~\cite{demler_superconducting_1997}
For inhomogeneous magnetization,
odd-frequency triplet correlations $f_{t1}$ with nonzero ($\pm 1$) total spin
projection are present as well. These correlations, not suppressed by the
exchange interaction, are long-ranged and have a dramatic impact
on transport properties and the Josephson effect.~\cite{bergeret_odd_2005}

Dominant influence of long-range triplet correlations on the Josephson current can be
realized in SFS junctions with magnetically active
interfaces,~\cite{eschrig_theory_2003,eschrig_triplet_2008}
narrow domain walls between S and thick F interlayers with misaligned
magnetizations,~\cite{braude_fully_2007,houzet_long_2007,volkov_odd_2008,
volkov_odd_2010,trifunovic_long-range_2010,alidoust_spin-triplet_2010-1} or
superconductors with spin orbit interaction.~\cite{PhysRevB.83.054513} In
general, fully developed triplet proximity effect can be realized only after inserting a
singlet-to-triplet ``converter," a thin (weak) ferromagnetic layer sandwiched
between a superconductor and a thick (strong) ferromagnet, acting as a ``filter''
which suppresses the short-range
correlations.\cite{braude_fully_2007,houzet_long_2007,volkov_odd_2008,%
volkov_odd_2010,trifunovic_long-range_2010}

The simplest superconductor-ferromagnet heterostructures with inhomogeneous
magnetization are F$_1$SF$_2$ and SF$_1$F$_2$S junctions with monodomain
ferromagnetic layers having noncollinear in-plane magnetizations. These
structures have been studied using Bogoliubov-de Gennes
equation~\cite{pajovi_josephson_2006,asano_josephson_2007,halterman_odd_2007,halterman_induced_2008}
and within the quasiclassical approximation in diffusive~\cite{fominov_triplet_2003,
you_magnetization-orientation_2004,loefwander_interplay_2005,barash_josephson_2002,fominov_josephson_2007,
crouzy_josephson_2007, sperstad_josephson_2008,karminskaya_josephson_2009} and
clean~\cite{blanter_supercurrent_2004,linder_proximity_2009} limits using Usadel
and Elienberger equations, respectively. It has been expected that the critical
supercurrent in SF$_1$F$_2$S junctions has a nonmonotonic dependence on angle
between magnetizations due to the long-ranged spin-triplet correlations.
However, in symmetric junctions a
monotonic dependence has been found both in the clean and dirty limits,
except for nonmonotonicity caused by $0-\pi$
transitions.~\cite{pajovi_josephson_2006,crouzy_josephson_2007}

Apparently, the long-range Josephson effect is not feasible in the
junctions with only two F layers,\cite{houzet_long_2007,volkov_odd_2010}
except in highly asymmetric
SF$_1$F$_2$S junctions at low temperatures, as we will
show here. In this case, the long-range spin-triplet
effect manifests itself as a large second harmonic ($I_2\gg I_1$)
in the spectral decomposition of the Josephson current-phase
relation, $I(\phi)=I_1 \sin(\phi)+I_2 \sin(2\phi)+\cdots$.
The ground state in Josephson junctions with ferromagnet can be $0$ or $\pi$
state.~\cite{buzdin_proximity_2005}
The energy of the junction is proportional to
$\int_0^{\phi}{I(\phi^{'})d\phi^{'}}$, hence a second harmonic leads to
degenerate ground states at $\phi=0$ and $\phi=\pi$.
Small contribution of the first harmonic lifts the degeneracy
which results in coexistence of stable and metastable $0$ and
$\pi$ states.\cite{radovi_coexistence_2001}

In this article, we study the Josephson effect and influence of odd-frequency
spin-triplet superconducting correlations in clean and moderately disordered
SF$_1$F$_2$S junctions with transparent interfaces. Magnetic interlayer is composed of two
monodomain ferromagnets with arbitrary orientation of in-plane magnetizations. We
calculate pair amplitudes and the Josephson current from the self-consistent
solutions of the Eilenberger equations. Pair amplitudes $f_s$ and $f_{t0}$ are
short-ranged, while $f_{t1}$ is long-ranged.

We show that the influence of misalignment of magnetizations on the Josephson current
in symmetric SFFS junctions with equal ferromagnetic layers
cannot be attributed to the emergence of long-range spin-triplet correlations:
For thin F layers (compared to the ferromagnetic exchange length) all pair amplitudes are
equally large, while for thick F layers the long-range triplet component
is very small. This explains why no substantial impact of long-range
spin-triplet superconducting correlations on the Josephson current has been
found previously for symmetric SFFS junctions both in the ballistic and
diffusive regimes.~\cite{pajovi_josephson_2006,crouzy_josephson_2007}

We find that long-range spin-triplet Josephson current can be realized only in
highly asymmetric SF$_1$F$_2$S junctions composed of particularly thin (weak) and thick (strong)
ferromagnetic layers with noncollinear magnetizations at low temperatures.
In that case the second harmonic in the Josephson current-phase relation is
dominant.
%No long-range spin-triplet current has been found for the
%ferromagnetic bilayer structure at temperatures close to $T_c$, in agreement with
%the results of Ref. \onlinecite{houzet_long_2007}.

The dominant second harmonic provides
more sensitive quantum interferometers (SQUIDs)
with effectively two times smaller flux
quantum\cite{radovi_coexistence_2001}
and gives the half-integer Shapiro steps.~\cite{sellier_half-integer_2004} The
coexistence of $0$ and $\pi$ ground-state configurations in
SQUIDs is potentially useful for
experimental study of the quantum superposition of macroscopically distinct
states.~\cite{friedman_quantum_2000,van_der_wal_quantum_2000}
Particularly asymmetric SF$_1$F$_2$S junctions also enable a robust realization of so-called
$\phi$-junctions.~\cite{buzdin_periodic_2003}
In general, Josephson junctions with a
non-sinusoidal current-phase relation are shown to be promising for realization
of ``silent'' phase qubits.~\cite{yamashita_superconducting_2005,orlando_superconducting_1999}

The reminder of the article is organized as follows. In Sec. II, we present the
model and the equations that we use to calculate the Josephson current and
spin-singlet and triplet pair amplitudes. In Sec. III, we provide numerical
results for different ferromagnetic layers thickness and strength and
orientation of magnetizations. The conclusion is given in Sec. IV.

\section{Model and formalism}

\begin{figure}
  \includegraphics[width=5.5cm]{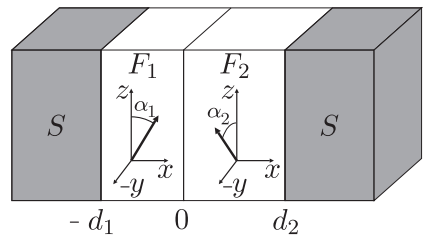}
  \caption{Schematics of an SF$_1$F$_2$S heterojunction. The magnetization
  vectors lie in the y-z plane at angles $\alpha_1$ and $\alpha_2$ with respect
  to the $z$-axis.}
  \label{geometrija}
\end{figure}

We consider a simple model of an SF$_1$F$_2$S  heterojunction consisting of two
conventional ($s$-wave, spin-singlet pairing) superconductors (S) and two uniform monodomain
ferromagnetic layers (F$_1$ and F$_2$) of thickness $d_1$ and $d_2$, with angle
$\alpha=\alpha_2-\alpha_1$ between their in-plane magnetizations (see
Fig.~\ref{geometrija}). Interfaces between layers are fully transparent and
magnetically inactive.

Superconductivity is described in the framework of the Eilenberger
quasiclassical theory.~\cite{bergeret_odd_2005,eilenberger_general_1966} Ferromagnetism
is modeled by the Stoner model, using an exchange energy shift $2h$ between the
spin subbands. Disorder is characterized by the electron mean free path
$l=v_F\tau$, where $\tau$ is the average time between  scattering on impurities,
and $v_F$ is the Fermi velocity assumed to be the same everywhere.

Both the clean and moderately diffusive ferromagnetic layers are considered.
In the clean limit, the mean free path $l$ is larger than the two characteristic
lengths: the ferromagnetic exchange length $\xi_F=\hbar v_F/h$, and the
superconducting coherence length $\xi_S=\hbar v_F/\pi\Delta_0$, where $\Delta_0$
is the bulk superconducting pair potential. For moderate disorder
$\xi_F<l<\xi_S$.

In this model, the Eilenberger Green functions
$g_{\sigma\sigma^{'}}(x,v_x,\omega_n)$,
$g^{\dag}_{\sigma\sigma^{'}}(x,v_x,\omega_n)$,
$f_{\sigma\sigma^{'}}(x,v_x,\omega_n)$, and
$f^{\dag}_{\sigma\sigma^{'}}(x,v_x,\omega_n)$ depend on the Cooper pair center-of-mass
coordinate $x$ along the junction axis, angle $\theta$ of the quasiclassical
trajectories with respect to the $x$ axis,
projection $v_x=v_F\cos{\theta}$ of the Fermi velocity vector,
and on the Matsubara frequencies $\omega_n=\pi k_BT(2n+1)$,
$n=0,\pm1,\ldots$. Spin indices are $\sigma=\uparrow,\downarrow$.

The Eilenberger equation in particle-hole $\otimes$ spin space can be written in
the compact form
\begin{equation}
  \label{Eil}
  \hbar v_x\partial_x \check{g}+\Big[\omega_n \hat{\tau}_3\otimes \hat{1}-i
  \check{V}-\check{\Delta}+\hbar \langle \check{g}\rangle/2\tau,
  \check{g}\Big]=0,
\end{equation}
with normalization condition $\check{g}^2=\check{1}$. We indicate by
$\hat{\cdots}$ and $\check{\cdots}$ $2\times2$ and $4\times4$ matrices,
respectively. The brackets $\langle\ldots\rangle$ denote angular averaging over
the Fermi surface (integration over $\theta$), and $[\;,\;]$ denotes a
commutator. The quasiclassical Green functions
\begin{equation}
  \check{g}=\left[
   \begin{array}{rrrr}
      g_{\uparrow\uparrow} & g_{\uparrow\downarrow} &
      f_{\uparrow\uparrow} & f_{\uparrow\downarrow}\\ g_{\downarrow\uparrow} &
      g_{\downarrow\downarrow} & f_{\downarrow\uparrow} &
      f_{\downarrow\downarrow}\\ -f_{\uparrow\uparrow}^{\dag} &
      -f_{\uparrow\downarrow}^{\dag} & -g_{\uparrow\uparrow}^{\dag} &
      -g_{\uparrow\downarrow}^{\dag}\\ -f_{\downarrow\uparrow}^{\dag} &
      -f_{\downarrow\downarrow}^{\dag} & -g_{\downarrow\uparrow}^{\dag} &
      -g_{\downarrow\downarrow}^{\dag}
    \end{array}  \right]
\end{equation}
are related to the corresponding Gor'kov-Nambu Green functions
$\check{G}=-i\langle T(\Psi\Psi^{\dag}) \rangle$
integrated over energy,
\begin{equation}
  \check{g}=\frac{i}{\pi}\hat{\tau}_3\otimes \hat{1}\int
  d\varepsilon_\mathbf{k}\vspace{2mm}\check{G},
\end{equation}
where
$\Psi=(\psi_{\uparrow},\psi_{\downarrow},
\psi^{\dag}_{\uparrow},\psi^{\dag}_{\downarrow})^{T}$
and $\varepsilon_{\mathbf{k}}=\hbar^2\mathbf{k}^2/2m-\mu$.
The matrix $\check{V}$ is given by
\begin{eqnarray}
  \check{V}=\hat{1}\otimes \textrm{Re}\Big[
  \textbf{h}(x)\cdot\widehat{\bm{\sigma}}\Big]+i\hat{\tau}_3\otimes
  \textrm{Im}\Big[ \textbf{h}(x)\cdot\widehat{\bm{\sigma}}\Big],
\end{eqnarray}
where the components  $\hat{\sigma}_x, \hat{\sigma}_y, \hat{\sigma}_z$ of the
vector $\widehat{\bm{\sigma}}$, and $\hat{\tau}_1, \hat{\tau}_2, \hat{\tau}_3$
are the Pauli matrices in the spin and the particle-hole space, respectively.
The in-plane ($y$-$z$) magnetizations of the neighboring F layers are not
collinear in general, and form angles $\alpha_1$ and $\alpha_2$ with respect to
the $z$-axis in the left (F$_1$) and the right (F$_2$) ferromagnets.
The exchange field in ferromagnetic layers is
$\textbf{h}(x)=h_{1}(0,\sin{\alpha_1},\cos{\alpha_1})$ and
$h_{2}(0,\sin{\alpha_2},\cos{\alpha_2})$.

We assume the superconductors are identical, with
\begin{equation}
  \check{\Delta}=\left[ \begin{array}{cc} 0 & \hat{\sigma}_2\Delta\\
    \hat{\sigma}_2\Delta^{*} & 0 \end{array}  \right]
\end{equation}
for $x<-d_1$ and $x>d_2$.
The self-consistency condition for the pair potential
$\Delta=\Delta(x)$ is given by
\begin{eqnarray}
  \label{samo}
  \Delta=-i\lambda 2\pi N(0) k_BT\sum_{\omega_n}\langle
  f_{\uparrow\downarrow}\rangle,
\end{eqnarray}
where $\lambda$ is the coupling constant, $N(0)=mk_F/2\pi^2\hbar^2$ is the
density of states per spin projection at the Fermi level
$E_F=\hbar^2 k_F^2/2m$, and $k_F=mv_F/\hbar$ is
the Fermi wave number. In F layers $\check{\Delta}=0$.

The supercurrent is obtained from the normal Green function through the following
expression
\begin{equation}
  \label{struja}
  I(\phi)=\pi e N(0)Sk_BT\sum_{\omega_n}\sum_{\sigma}\langle v_x
  \textrm{Im}\hspace{1mm} g_{\sigma\sigma}\rangle ,
\end{equation}
where $\phi$ is the macroscopic phase difference across the junction, and $S$ is
the area of the junction. In examples, the current is normalized to the
resistance $R_N=2\pi^2\hbar/Se^2k_F^2$.

Pair amplitudes, singlet $f_s$, and triplet $f_{t0}$ and $f_{t1}$, with  0 and
$\pm1$ projections of the total spin of a pair, are defined in
terms of anomalous Green functions
\begin{eqnarray} f_s(x,t)&=&-i\pi N(0) k_BT\sum_{\omega_n}\langle
  f_{\uparrow\downarrow}-f_{\downarrow\uparrow}\rangle e^{-i \omega_n t},\\
  f_{t0}(x,t)&=& -i\pi N(0) k_BT\sum_{\omega_n}\langle
  f_{\uparrow\downarrow}+f_{\downarrow\uparrow}\rangle e^{-i \omega_n t},\\
  f_{t1}(x,t)&=&-i\pi N(0) k_BT\sum_{\omega_n}\langle
  f_{\uparrow\uparrow}+f_{\downarrow\downarrow}\rangle e^{-i \omega_n t}.
\end{eqnarray}

In the following we will characterize singlet amplitudes by the zero-time
$f_s=f_s(x,0)$. However, since zero-time triplet amplitudes identically vanish
in agreement with the Pauli principle
(with or without self-consistency),\cite{halterman_odd_2007} we
characterize triplets by auxiliary functions using summation over negative
frequencies only,
\begin{eqnarray}
  f_{t0}^{<}&=& -i\pi N(0) k_BT\sum_{\omega_n<0}\langle
  f_{\uparrow\downarrow}+f_{\downarrow\uparrow}\rangle,\\ f_{t1}^{<}&=&-i\pi
  N(0) k_BT\sum_{\omega_n<0}\langle
  f_{\uparrow\uparrow}+f_{\downarrow\downarrow}\rangle.
\end{eqnarray}

Note that in previous definitions of triplet pair amplitudes the total spin of a pair
is projected on the $z$-axis. Physically it is more reasonable to take
directions of magnetizations in F layers as the spin quantization axes.
New triplet amplitudes can be introduced in the F$_1$ layer
($-d_1<x<0$) by simple rotation
\begin{eqnarray} \label{t1}
-i\tilde{f}_{t0}
&=&
\cos(\alpha_1)(-if_{t0}) - \sin(\alpha_1)f_{t1},
\\
\tilde{f}_{t1}
&=&
\sin(\alpha_1)(-if_{t0}) + \cos(\alpha_1)f_{t1}. \label{f2}
\end{eqnarray}
The same expressions with $\alpha_1\to \alpha_2$ hold for F$_2$
layer, $0\leq x< d_2$. Auxiliary functions $\tilde{f}_{t0}^{<}$ and
$\tilde{f}_{t1}^{<}$ are related to $f_{t0}^{<}$ and $f_{t1}^{<}$ in the same
way.

For $\phi=0$,  amplitudes $f_s$ and
$f_{t1}^{<}$ are real, while $f_{t0}^{<}$ is imaginary, and for $\phi=\pi$  the
opposite is true.  For $0<\phi<\pi$, all amplitudes are complex.
In the examples, we normalize the amplitudes to the value of $f_s$ in
the bulk superconductors,

\begin{equation}
  f_{sb}=2\pi N(0)
  k_BT\sum_{\omega_n}\frac{\Delta_0}{\sqrt{\omega_n^2+\Delta_0^2}}.
\end{equation}
Here, in the summation over $\omega_n$, the high-frequency cutoff of $10\Delta_0$ is
used. The temperature dependence of the bulk pair potential $\Delta_0 $ is given
by $\Delta_0 (T)=\Delta_0 (0)\tanh \left(
1.74\sqrt{T_{c}/T-1}\right)$.~\cite{muehlschlegel_thermodynamischen_1959}

We consider only transparent interfaces and use continuity of the Green
functions as the boundary conditions.  For planar SF$_1$F$_2$S junctions (three
dimensional case)  Eq.~(\ref{Eil}) is solved  iteratively together with the
self-consistency condition, Eq.~(\ref{samo}). The averaging over
$\theta$ is given by $\langle ...\rangle=(1/2)\int_0^{\pi}(...)d(\cos \theta)$.

Numerical computation is carried out using the collocation method.
Iterations are performed until self-consistency is reached, starting from the
stepwise approximation for the pair potential
\begin{equation}
  \label{del}
  \Delta=\Delta_0 \left[ e^{-i\phi /2}\Theta
  (-x-d_{1})+e^{i\phi /2}\Theta (x-d_{2})\right],
\end{equation}
where $\Theta(x)$ is the Heaviside step function. For finite electron mean free
path in ferromagnets, the iterative procedure starts from the clean limit. We
choose the appropriate boundary conditions in
superconductors at the distance exceeding $2\xi_S$ from the SF interfaces.
These boundary conditions are given by eliminating the unknown constants from the
analytical solutions in stepwise approximation. To reach self-consistency
starting from the stepwise $\Delta$, five to ten iterative steps were sufficient
for results shown in this article.

\section{Results}

\begin{figure}
  \includegraphics[width=8cm]{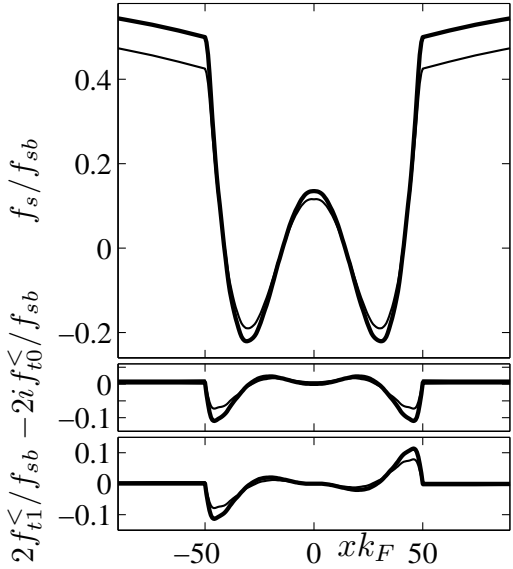}
  \caption{Spatial dependence of singlet and triplet pair amplitudes $f_s$,
  $f_{t0}^{<}$, and $f_{t1}^{<}$,  normalized to the bulk singlet amplitude
  $f_{sb}$. We consider a symmetric SF$_1$F$_2$S junction ($d_1=d_2=50k_F^{-1}$)
  at low temperature, $T/T_c=0.1$, with the electron mean free path in
  ferromagnets $l=200k_F^{-1}$ (thick curves) and in the clean limit
  $l\to \infty$ (thin curves). The phase difference is
  $\phi=0$. The magnetizations in ferromagnets are orthogonal,
  $\alpha_1=-\pi/4$, $\alpha_2=\pi/4$, and $h_{1}=h_{2}=0.1E_F$.
  Superconductor is characterized by $\Delta_0(0) /E_{\rm F}=10^{-3}$.}
  \label{f50}
\end{figure}

We illustrate our results on SF$_1$F$_2$S planar junctions with relatively weak
ferromagnets, $h/E_F\sim 0.1$, and the ferromagnetic exchange length $\xi_F\sim
20k_F^{-1}$. Superconductors are characterized by the bulk pair potential at
zero temperature $\Delta_0(0) /E_{\rm F}=10^{-3}$, which corresponds to the
superconducting coherence length $\xi_S(0)=636 k_F^{-1}$.
We assume that all interfaces are fully
transparent and the Fermi wave numbers in all metals are equal ($k_F^{-1}\sim$ 1{\AA}).

\begin{figure}
  \includegraphics[width=8cm]{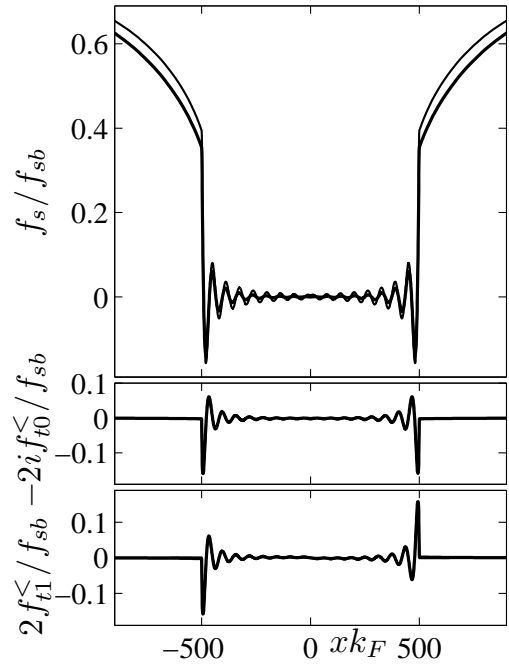}
  \caption{The same as in Fig.~\ref{f50} for ten times thicker ferromagnetic
  films, $d_1=d_2=500k_F^{-1}$.}
  \label{f500}
\end{figure}

\begin{figure}
  \includegraphics[width=8cm]{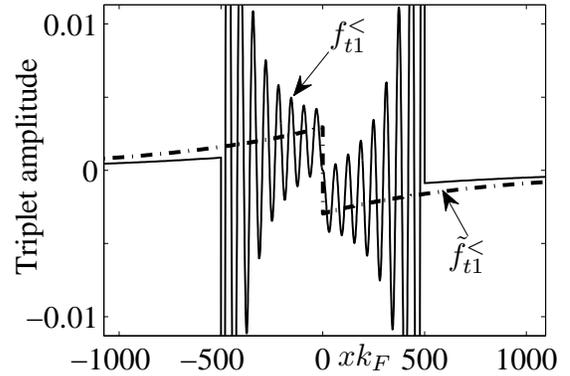}
  \caption{Comparison between normalized triplet amplitudes $2 f_{t1}^</f_{sb}$
  (solid curve) and rotated $2 \tilde{f}_{t1}^</f_{sb}$ (dash-dotted curve).
  Parameters are the same as in Fig.~\ref{f500} for $l\to \infty$.} \label{frot}
\end{figure}

\begin{figure}
  \includegraphics[width=8cm]{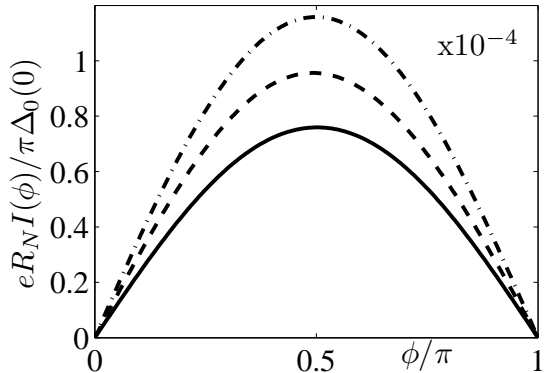} \caption{The
  current-phase relation $I(\phi)$ for a symmetric SF$_1$F$_2$S junction with
  $d_1=d_2=500k_F^{-1}$, $h_{1}=h_{2}=0.1E_F$, $l=200k_F^{-1}$, $T/T_c=0.1$,
  and for three values of the relative angle between magnetizations:
  $\alpha=0$ (solid curve), $\pi/2$ (dashed curve), and $\pi$ (dash-dotted
  curve).}
  \label{I500}
\end{figure}

\begin{figure}
  \includegraphics[width=8cm]{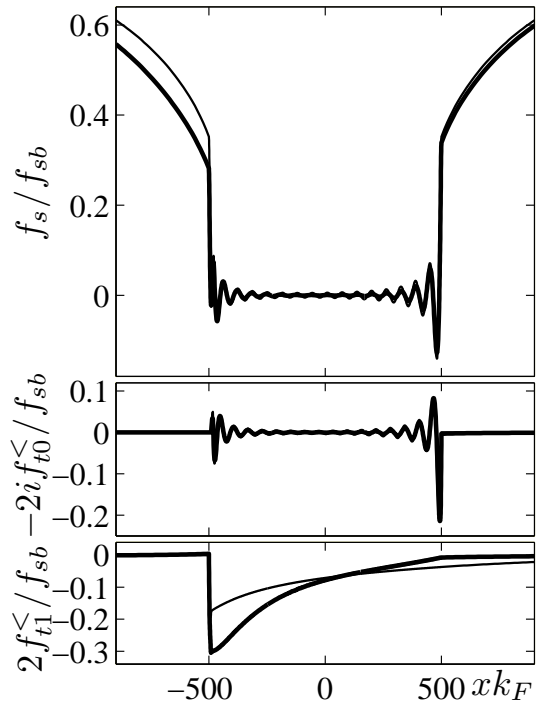}
  \caption{Spatial dependence of singlet and triplet pair amplitudes $f_s$,
  $f_{t0}^{<}$, and $f_{t1}^{<}$,  normalized to the bulk singlet amplitude
  $f_{sb}$. We consider an asymmetric SF$_1$F$_2$S junction ($d_1=10k_F^{-1}$ and
  $d_2=990k_F^{-1}$) at low temperature, $T/T_c=0.1$, with the electron mean
  free path in ferromagnets $l=200k_F^{-1}$ (thick curves) and in the clean
  limit $l\to \infty$ (thin curves). The phase difference is
  $\phi=0$. Magnetization in the thin layer is along $y$-axis,
  $\alpha_1=-\pi/2$, and along $z$-axis in the thick layer, $\alpha_2=0$. Here,
  $h_1=h_2=0.1E_F$ and $\Delta_0(0) /E_{\rm F}=10^{-3}$.}
  \label{f500asim}
\end{figure}

\begin{figure}
  \includegraphics[width=7.9cm]{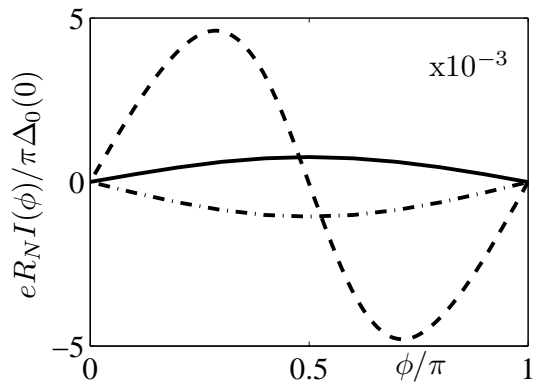}
  \caption{The current-phase relation $I(\phi)$ for an asymmetric SF$_1$F$_2$S
  junction with $d_1=10k_F^{-1}$ and $d_2=990k_F^{-1}$, $h_1=h_2=0.1E_F$,
  $l=200k_F^{-1}$, $T/T_c=0.1$, and for three values of the relative angle
  between magnetizations: $\alpha=0$ (solid curve), $\pi/2$ (dashed curve), and
  $\pi$ (dash-dotted curve).}
  \label{I500asim}
\end{figure}

\begin{figure}
  \includegraphics[width=7.9cm]{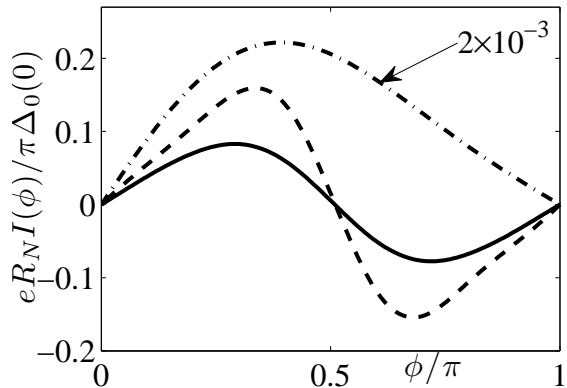}
  \caption{The current-phase relation $I(\phi)$ for an asymmetric SF$_1$F$_2$S
  junction in the clean limit $l\rightarrow \infty$ with orthogonal
  magnetizations $\alpha=\pi/2$. For $d_1=10k_F^{-1}$, $d_2=990k_F^{-1}$,
  $h_1=h_2=0.1E_F$ three cases are shown: $T=0.1T_c$ (thick solid curve);
  $T=0.01T_c$ (dashed curve); $T=0.9T_c$ (dash-dotted curve, magnified $5\times
  10^2$ times).}
  \label{Tdependence}
\end{figure}

\begin{figure}
  \includegraphics[width=8cm]{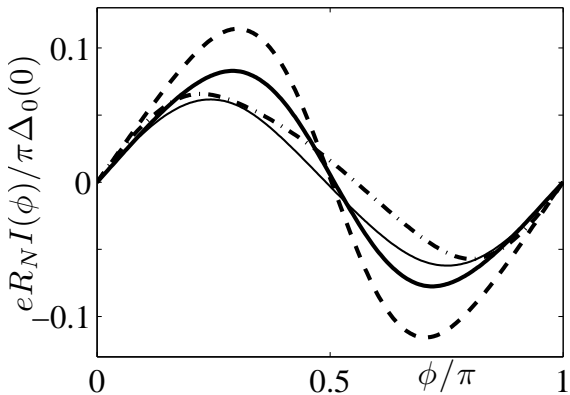}
  \caption{The current-phase relation $I(\phi)$ for an asymmetric SF$_1$F$_2$S
  junction in the clean limit $l\rightarrow \infty$ with orthogonal
  magnetizations $\alpha=\pi/2$, and $T/T_c=0.1$. For $d_1=10k_F^{-1}$,
  $d_2=990k_F^{-1}$, three cases are shown: $h_1=h_2=0.1E_F$ (thick solid
  curve); $h_1=0.1E_F$, $h_2=0.2E_F$ (dashed curve); $h_1=0.05E_F$, $h_2=0.1E_F$
  (dash-dotted curve). Influence of F layer thickness is shown for
  $d_1=20k_F^{-1}$, $d_2=980k_F^{-1}$, and $h_1=h_2=0.1E_F$ (thin solid curve).}
  \label{poslednja}
\end{figure}

Detailed analysis is given for symmetric junctions with relatively thin
($d_{1}=d_{2}=50k_F^{-1}$) and thick ($d_{1}=d_{2}=500k_F^{-1}$) ferromagnetic
layers, Figs.~\ref{f50}--\ref{I500}, and for an highly asymmetric junction
($d_1=10k_F^{-1}$ and $d_2=990k_F^{-1}$), Figs.~\ref{f500asim} and
\ref{I500asim}. In these examples $h_1=h_2=0.1E_F$, $T/T_c=0.1$, and both the
clean limit ($l\rightarrow\infty$) and moderate disorder in ferromagnets
($l=200k_{F}^{-1}$) are considered. The influences of temperature, thickness and
exchange field variations are shown in Figs.~\ref{Tdependence} and
\ref{poslednja} in the clean limit.

Short-ranged pair amplitudes $f_s$ and $f_{t0}^{<}$ decay spatially from the FS
interfaces in the same oscillatory manner. They decay algebraically with
length $\hbar v_F/h$ in the clean limit, and exponentially with the characteristic length
$\sqrt{\hbar D/h}$ in the dirty limit, where diffusion coefficient $D=v_Fl/3$.

The long-ranged pair amplitude $f_{t1}^{<}$ is not suppressed by the exchange field and
penetrates the ferromagnet on the scale $\hbar v_F/k_BT$ ($\sqrt{\hbar
D/k_BT}$) in the clean (dirty) limit. In symmetric junctions ($d_1=d_2$,
$h_1=h_2$, and $\alpha_1=-\alpha/2$, $\alpha_2=\alpha/2$) all pair
amplitudes are practically the same in the clean limit ($l\to
\infty$) and for moderate disorder ($l=200k_F^{-1}$) in ferromagnets, which is
illustrated in Figs.~\ref{f50} and \ref{f500}. In symmetric junctions,
$f_{t1}^<$ does not prevail; it can be seen in Fig. \ref{I500} that
$\alpha$-dependence of the Josephson critical
current is monotonic.

Note that the spatial oscillations of $f_{t1}^{<}$ in Figs. \ref{f50}-\ref{frot} are
due to our choice of $z$ axis as the spin quantization axis.
After choosing the magnetization in F layers as the spin
quantization axis, the rotated amplitude $\tilde{f}_{t1}^{<}$ no
longer oscillates but instead exhibits monotonic spatial variation
with the jump at $x=0$ (dash-dotted curve in
Fig.~\ref{frot}). It can be seen that the amplitude  $\tilde{f}_{t1}^{<}$ decays
from F$_1$F$_2$ interface.  This is because $\tilde{f}_{t1}^{<}$ in each F
layer is generated at the interface itself by the projection
$\tilde{f}_{t0}^{<}\sin \alpha$ from the neighboring layer.
The triplet pair amplitudes penetrate into superconductors and monotonically decay over
the same distance (the bulk superconducting coherence length $\xi_S$) as the
singlet amplitude saturates to the bulk value (see Fig. \ref{frot}).

For thick F layers (when short-ranged amplitude $\tilde{f}_{t0}^{<}$ is highly suppressed
at the F$_1$F$_2$ interface) the generated long-ranged component
$\tilde{f}_{t1}^{<}$ is very small. Therefore, influence of misalignment of
magnetizations on the Josephson current in symmetric SF$_1$F$_1$S junctions
cannot be attributed to the emergence of spin-triplet correlations: For thin
ferromagnetic layers all amplitudes are equally large, while for thick
ferromagnetic layers the long range triplet $\tilde{f}_{t1}$ is very small.

Fully developed long-range spin-triplet proximity effect emerges in
asymmetric junctions with particularly
thin and thick F layers (or weak and strong ferromagnets). The thin (weak) F layer acts as a
``triplet-generator", while the thick (strong) F layer is a ``filter" which suppresses
the short-ranged components. This is illustrated in Fig.~\ref{f500asim} for
thin F$_1$ layer ($d_1=10k_F^{-1}$) and thick F$_2$ layer ($d_2=990k_F^{-1}$)
of equal strength $h_1=h_2=0.1\, E_F$,
with orthogonal magnetizations, $\alpha_1=-\pi/2$ and $\alpha_2=0$.
Here, magnetization in the thick (F$_2$) layer is taken along
the $z$-axis and along the $y$-axis in the thin (F$_1$) layer. Because of this
choice, the $x$-dependence of  $f_{t1}^<$ is monotonic in F$_2$. The
thin layer thickness $d_1$ is chosen to give maximum triplet current for a
moderate disorder in ferromagnets ($l=200k_F^{-1}$). This explains why
$f_{t1}^<$ in Fig.~\ref{f500asim} is larger for the finite electron mean-free
path than in the clean limit.

The current-phase relation for an highly asymmetric SF$_1$F$_2$S junction
(the same parameters as in Fig.~\ref{f500asim}) is shown in
Fig.~\ref{I500asim}.
The critical current for orthogonal magnetizations is an order of
magnitude larger than for collinear magnetizations even though the
first harmonic is absent. The second harmonic is dominant like at
$0$-$\pi$ transitions,\cite{radovi_coexistence_2001,radovi_josephson_2003}
except being larger in magnitude and long-ranged in the present case.

Robustness of the long-ranged second harmonic is illustrated
in Figs.  \ref{Tdependence} and \ref{poslednja} in the clean limit.
The magnitude of the second harmonic is enhanced by lowering the
temperature, as can be seen from Fig. \ref{Tdependence}.
This is in contrast to the temperature-induced $0$-$\pi$ transition in
which the first harmonic is restored away from the transition
temperature.\cite{radovi_josephson_2003,RyazanovPRL86-01,KontosApriliPRL89-02}
When the temperature is close to the critical
temperature $T_c$, the triplet proximity effect, i.e. the second harmonic in $I(\phi)$, is no longer dominant in an agreement with the results of Ref.
\onlinecite{houzet_long_2007}. Small variations of the layers thickness and the
exchange energy of ferromagnets (Fig. \ref{poslednja}), as well as moderate
disorder (Fig. \ref{I500asim}), do not affect the dominance of the long-ranged second
harmonic at low temperatures.
Since the influence of ferromagnet is determined by
$\Theta=(h/E_F)k_F d$, the shown results remain the same for
$\Theta_1\sim 1$ in F$_1$ and $\Theta_2\sim 100$ in F$_2$ layers.

In all calculations we have assumed
transparent SF interfaces and clean or moderately disordered ferromagnets.
Finite transparency much strongly suppresses higher harmonics than
the first one and the considered effect becomes negligible in the
tunnel limit. We expect the similar influence of large disorder:
In diffusive SFS junctions
at $0$-$\pi$ transition,\cite{RyazanovPRL86-01,KontosApriliPRL89-02}
where the second harmonic is dominant,
the Josephson current drops to zero. However, this is not the case
in the clean SFS junctions where the Josephson current is smaller but
finite at the $0$-$\pi$ transition.\cite{radovi_josephson_2003}

\section{Conclusion}

We have studied the Josephson effect in SF$_1$F$_2$S planar
junctions made of conventional superconductors and two
monodomain ferromagnetic layers with arbitrary thickness, strength,
and angle between in-plane magnetizations. We carried out a detailed
analysis of spin-singlet and -triplet pairing correlations and the
Josephson current in the clean limit and for moderate disorder in
ferromagnets by solving self-consistently the Eilenberger
equations. While the spin-singlet and -triplet correlations with
zero spin projection are short range, the triplet
correlations with a nonzero spin projection are long range and may
have a dramatic impact on transport properties and the Josephson
effect.

In symmetric SFFS junctions with ferromagnetic layers of
equal strength,
the long-range spin-triplet correlations have no substantial impact
on the Josephson current both in the ballistic and moderately
diffusive regimes: For thin (weak) ferromagnetic layers all amplitudes
are equally large, while for thick (strong) layers the long-range triplet
amplitude is very small. This explains the previous
results of Refs.~\onlinecite{pajovi_josephson_2006},~\onlinecite{crouzy_josephson_2007}.

We have found that fully developed long-range spin-triplet
proximity effect occurs in highly asymmetric SF$_1$F$_2$S
junctions at low temperatures and manifests itself as a dominant
second harmonic in the Josephson current-phase relation. In
contrast to the temperature-induced $0$-$\pi$ transition in which
the first harmonic is restored away from the transition temperature,
the magnitude of the second harmonic due to long-range
spin-triplet correlations increases as the temperature is lowered.
The triplet-induced second harmonic is robust against
moderate disorder and variations in the layers thickness and
exchange energy of ferromagnets.

Dominant second harmonic and the resulting ground state degeneracy
of the Josephson junction (like at $0$-$\pi$ transitions) is
experimentally accessible in asymmetric SF$_1$F$_2$S junctions
with small interface roughness at low temperatures.
In addition, the half-periodicity of $I(\phi)$ in the considered junctions can
be used for quantum interferometers (SQUIDs)
which operate with two times smaller flux quantum and
can exhibit the superposition of macroscopically distinct quantum
states even in the absence of an external magnetic
field.~\cite{radovi_coexistence_2001}
This has a potential
application in the field of quantum
computing.\cite{orlando_superconducting_1999}

\vspace*{-5mm}
\section{Acknowledgment}
\vspace*{-5mm}

We acknowledge Mihajlo Vanevi{\'c} for help and for valuable
discussions. The work was supported by the Serbian Ministry of
Science, Project No.~171027.
\bibliography{condmat}
\end{document}